\pdfoutput=1

\documentclass[12pt]{article}
\usepackage{graphicx}
\usepackage{subfig}
\usepackage{xspace}
\usepackage{multirow,booktabs}
\usepackage{amsmath,amsthm,amssymb}
\usepackage{url}
\usepackage{hyperref}

\def\etal{{\it et al.}\xspace}

\hypersetup{%
    pdftitle = {Average value of available measurements of the absolute air-fluorescence yield},
    pdfsubject = {GAP note},
    pdfkeywords = {Air-fluorescence yield, fluorescence telescopes},
    pdfauthor = {J. Rosado, F. Blanco, F. Arqueros},
    pdfstartview = {FitH},
    }
\begin{document}
\begin{titlepage}
\begin{center}
\Huge {\bf Average value of available measurements of the absolute air-fluorescence yield}
\par
\vspace*{2.0cm} \normalsize {\bf {J. Rosado, F. Blanco, F. Arqueros}}
\par
\vspace*{0.5cm} \small \emph{Departamento de F\'{i}sica At\'{o}mica, Molecular y Nuclear, Facultad de Ciencias
F\'{i}sicas, Universidad Complutense de Madrid, E-28040 Madrid, Spain}
\end{center}
\vspace*{2.0cm}

\begin{abstract}
The air-fluorescence yield is a key parameter for determining the energy scale of ultra-high-energy cosmic rays
detected by fluorescence telescopes. A compilation of the available measurements of the absolute air-fluorescence yield
normalized to its value in photons per MeV for the 337~nm band at given pressure and temperature has been recently
presented in~\cite{Rosado}. Also, in that paper, some corrections in the evaluation of the energy deposited in the
corresponding experimental collision chambers have been proposed. In this note this comparison is updated. In addition,
a simple statistical analysis is carried out showing that our corrections favor the compatibility among the various
experiments. As a result, an average value of 5.45~ph/MeV for the fluorescence yield of the 337~nm band (20.1~ph/MeV
for the spectral interval $300-420$~nm) at 1013~hPa and 293~K with an uncertainty of 5\% is found. This result is fully
compatible with that recently presented by the AIRFLY collaboration (still preliminary) in such a way that including
this latest result could even lowered the final uncertainty below the 5\% level with high reliability.
\end{abstract}

\end{titlepage}


The fluorescence technique has been proved to be very fruitful for the study of ultra-high-energy cosmic rays. The
air-fluorescence yield, defined as the number of fluorescence photons per unit deposited energy, is a key calibration
parameter of the fluorescence telescopes. In~\cite{Rosado}, we presented a comparison between the laboratory
measurements of the absolute air-fluorescence yield available in the literature normalized to common units
(photons/MeV), air conditions (dry air, 800~hPa and 293~K) and wavelength interval (337~nm band). From the comparison
of this normalized $Y_{337}$ parameter, discrepancies larger than the quoted uncertainties of the measurements were
found. According to the Monte Carlo analysis carried out in~\cite{Rosado}, non-negligible corrections in the evaluation
of the energy deposition should be applied to these results, in particular in experiments where authors neglected the
energy deposited outside the field of view of the light detector. In fact, the proposed corrections are even larger
than the quoted uncertainties in some cases (i.e., measurements of Kakimoto~\etal~\cite{Kakimoto} and
Lefeuvre~\etal~\cite{Lefeuvre}). As a result of these corrections, the $Y_{337}$ values get closer to each other in
such a way that they are contained within the $6-7$~ph/MeV interval at the above reference pressure and
temperature~\cite{Rosado}.

In this work, the calculations presented in~\cite{Rosado} have been updated and the compatibility of the fluorescence
yield results with and without our corrections has been evaluated quantitatively. This has allowed to obtain a reliable
$Y_{337}$ value resulting from an average of the available absolute measurements. With respect to the MC analysis
performed in~\cite{Rosado}, some improvements in the algorithm for the evaluation of the energy deposited in the
collision chamber of the experiments have been made. In particular, the density correction on the cross section for the
ionization processes leading to ejection of K-shell electrons, which was neglected in~\cite{Rosado}, have been included
in this work. These highly excited ions generate X rays which deposit all their energy within a very short distance. As
a consequence, the density correction for the K shell lowers the energy deposited by high-energy electrons (GeV range).
GEANT4 simulations of energy deposition in air carried out by MACFLY~\cite{MACFLY} and AIRFLY~\cite{GEANT4_AIRFLY} are
in agreement with our calculations at the level of 2\% (1\%) for electron energies in the GeV (MeV) range.

In table~\ref{tab:summary}, we present the compilation of fluorescence yield values~\cite{Rosado} with these updated
corrections\footnote{The updated corrections are very similar to those proposed in~\cite{Rosado} except for experiments
working with very-high-energy electrons; however, the average fluorescence yields presented below are basically not
affected by this update.}. In this case the fluorescence yield have been normalized to its value at 1013~hPa (instead
of 800~hPa as presented in~\cite{Rosado}) and 293~K. The latest result of AIRFLY~\cite{AIRFLY} carried out using
120~GeV protons have been included in this compilation, but the impact on the average has been studied separately, as
this value is still preliminary.

\begin{table}[t!]
\centering\small%
\caption{\footnotesize Comparison of available measurements of the absolute air-fluorescence yield normalized to the
337~nm band at 1013~hPa and 293~K (dry air). Experiments are listed in column~1, the electron energies are shown in
column~2 and the corresponding results, both uncorrected and corrected (in bold), are listed in column~3. The
experimental uncertainties quoted by the authors are shown in column~4 and the size of the proposed correction is
displayed in the last column.}
\begin{tabular}{*{5}{c}}
\addlinespace\toprule
Experiment                                &     $E$ (MeV)    & $Y_{337}$ (ph/MeV) &      Quoted error      & Correction \\
\midrule\midrule
\multirow{4}{*}{Kakimoto~\cite{Kakimoto}} &        1.4       & 4.5 / \textbf{4.8} & \multirow{4}{*}{10\%}  &     +6\%   \\
                                          &      300         & 4.4 / \textbf{5.5} &                        &    +25\%   \\
                                          &      650         & 3.8 / \textbf{4.8} &                        &    +27\%   \\
                                          &     1000         & 4.3 / \textbf{5.5} &                        &    +29\%   \\
\midrule
Nagano~\cite{Nagano_04}                   &        0.85      & 5.0 / \textbf{5.4} &         13\%           &     +6\%   \\
\midrule
\multirow{2}{*}{Lefeuvre~\cite{Lefeuvre}} &        1.1       & 5.1 / \textbf{5.5} &  \multirow{2}{*}{5\%}  &     +7\%   \\
                                          &        1.5       & 5.6 / \textbf{6.1} &                        &     +8\%   \\
\midrule
\multirow{3}{*}{MACFLY~\cite{MACFLY}}     &        1.5       & 4.3 / \textbf{4.4} & \multirow{3}{*}{13\%}  &     +1\%   \\
                                          &  $20\cdot 10^3$  & 4.4 / \textbf{4.3} &                        &     -2\%   \\
                                          &  $50\cdot 10^3$  & 4.6 / \textbf{4.5} &                        &     -2\%   \\
\midrule
FLASH~\cite{FLASH_08}                     & $28.5\cdot 10^3$ & 5.5 / \textbf{5.6} &          7.5\%         &     +2\%   \\
\midrule
AirLight~\cite{AirLight}                  &      $0.2-2$     & 5.8 / \textbf{5.4} &           16\%         &     -7\%   \\
\midrule
AIRFLY~\cite{AIRFLY}                      & $120\cdot 10^3$  & 5.6 / \textbf{ - } &      $\lesssim5\%$     &      -     \\
\bottomrule\addlinespace%
\end{tabular}\label{tab:summary}
\end{table}

\begin{figure}[t!]
\centering
\includegraphics[width=0.8\linewidth]{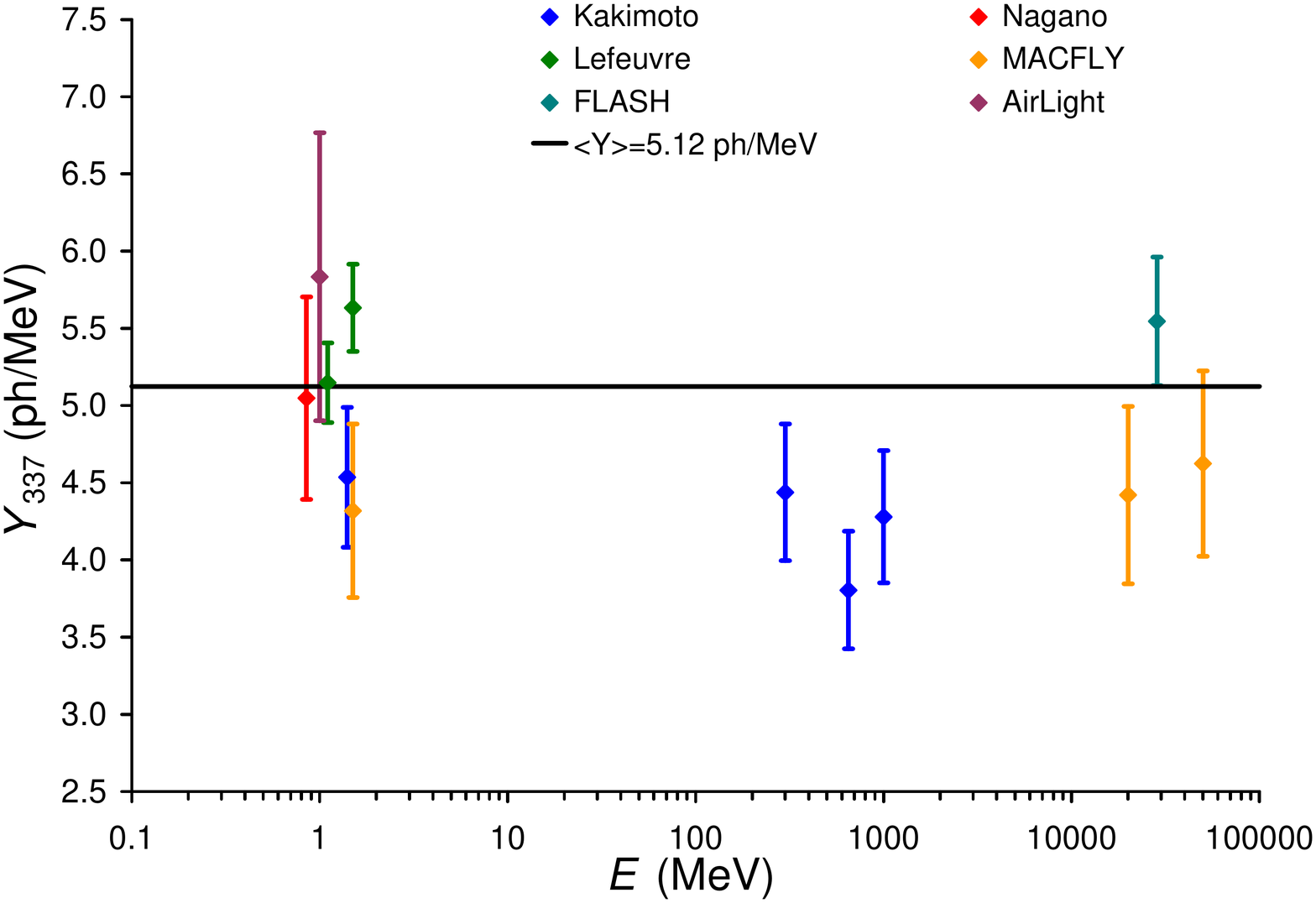}%
\caption{\footnotesize Comparison of uncorrected normalized $Y_{337}$ values as a function of energy. The horizontal
line represents the corresponding weighted average value (see text for details).}%
\label{fig:uncorr_E}%
\end{figure}

\begin{figure}[t!]
\centering
\includegraphics[width=0.8\linewidth]{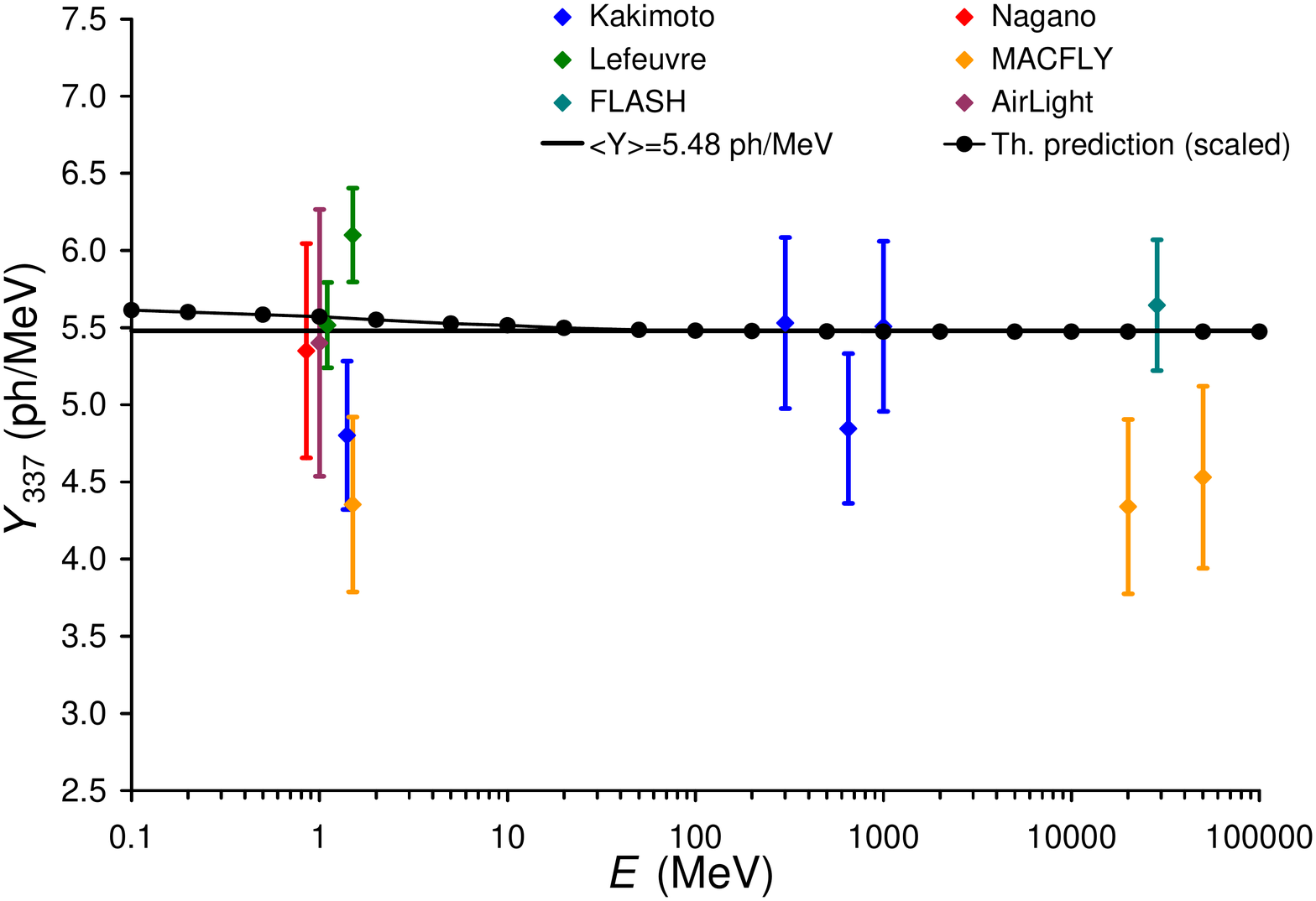}%
\caption{\footnotesize Same as figure~\ref{fig:uncorr_E} for the corrected $Y_{337}$ values. The black
circles connected by lines represent the weak energy dependence predicted by our simulation~\cite{Arqueros_NJP}.}%
\label{fig:corr_E}%
\end{figure}

The fluorescence yield results have been displayed as a function of energy in figures~\ref{fig:uncorr_E} (uncorrected
values) and~\ref{fig:corr_E} (corrected ones). As can be appreciated, they are in a better agreement when including our
corrections. In addition, the corrected results give more support to the expected energy independence of the
fluorescence yield.

In order to know quantitatively to what extent our corrections favor the agreement between the absolute results
included in this comparison, a statistical analysis has been performed. In the first place, for a given experiment,
results obtained at different energies have been averaged assuming a common systematic error\footnote{This way, no
higher significance is given to experiments measuring at several energies than those measuring at a single energy.}.
The results are shown in table~\ref{tab:average} for both the uncorrected (column~2) and corrected (column~3) $Y_{337}$
values together with the percentage uncertainty associated to each experiment (column~4).

\begin{table}[t!]
\centering\small%
\caption{\footnotesize Comparison of absolute $Y_{337}$ values of table~\ref{tab:summary}, where measurements reported
by a given experiment at different electron energies have been averaged.}
\begin{tabular}{*{4}{c}}
\addlinespace\toprule
Experiment               & $Y^{\rm uncorr}_{337}$ (ph/MeV) & $Y^{\rm corr}_{337}$ (ph/MeV) & Quoted error  \\
\midrule\midrule
Kakimoto~\cite{Kakimoto} &               4.3               &             5.2               &      10\%     \\
\midrule
Nagano~\cite{Nagano_04}  &               5.0               &             5.4               &      13\%     \\
\midrule
Lefeuvre~\cite{Lefeuvre} &               5.4               &             5.8               &       5\%     \\
\midrule
MACFLY~\cite{MACFLY}     &               4.5               &             4.4               &      13\%     \\
\midrule
FLASH~\cite{FLASH_08}    &               5.5               &             5.6               &      7.5\%    \\
\midrule
AirLight~\cite{AirLight} &               5.8               &             5.4               &      16\%     \\
\midrule
AIRFLY~\cite{AIRFLY}     &               5.6               &              -                & $\lesssim5\%$ \\
\bottomrule\addlinespace%
\end{tabular}\label{tab:average}
\end{table}

In the second place, the weighted average of the data sample shown in table~\ref{tab:average} has been computed from

\begin{equation}\label{mean}
\langle Y \rangle=\frac{\sum_i{w_i Y_i}}{\sum_i {w_i}}\,,
\end{equation}
using as weights the reciprocal of the quoted square uncertainties (i.e., $w_i=1/\sigma_i^2$). If these $\sigma_i^2$
are assumed to actually represent the variances of the corresponding (normal) probability distributions, our weighted
mean would be the best estimator of the $Y_{337}$ value with a variance given by $\left(\sum_i
1/\sigma_i^2\right)^{-1}$. However, note that authors usually do not include any error contribution from the evaluation
of the deposited energy, and thus, uncertainties are very likely underestimated, at least in some experiments. In fact,
the $\chi^2$ statistic divided by the number of degrees of freedom ($\chi^2/{\rm ndf}$) is found to be somewhat larger
than 1 for the uncorrected sample. Therefore, following the usual procedure, the variance of the weighted mean was
corrected multiplying by $\chi^2/{\rm ndf}$, that is,

\begin{equation}\label{sigma2_average}
\sigma^2_{\langle Y\rangle}=\frac{\chi^2/{\rm ndf}}{\sum_i 1/\sigma_i^2}=
\frac{\sum_i{w_i\left(Y_i-\langle Y\rangle\right)^2}}{(n-1)\sum_i {w_i}}\,.
\end{equation}

\begin{figure}[t!]
\centering
\subfloat[\footnotesize Uncorrected $Y_{337}$ values.]{%
\includegraphics[width=0.7\linewidth]{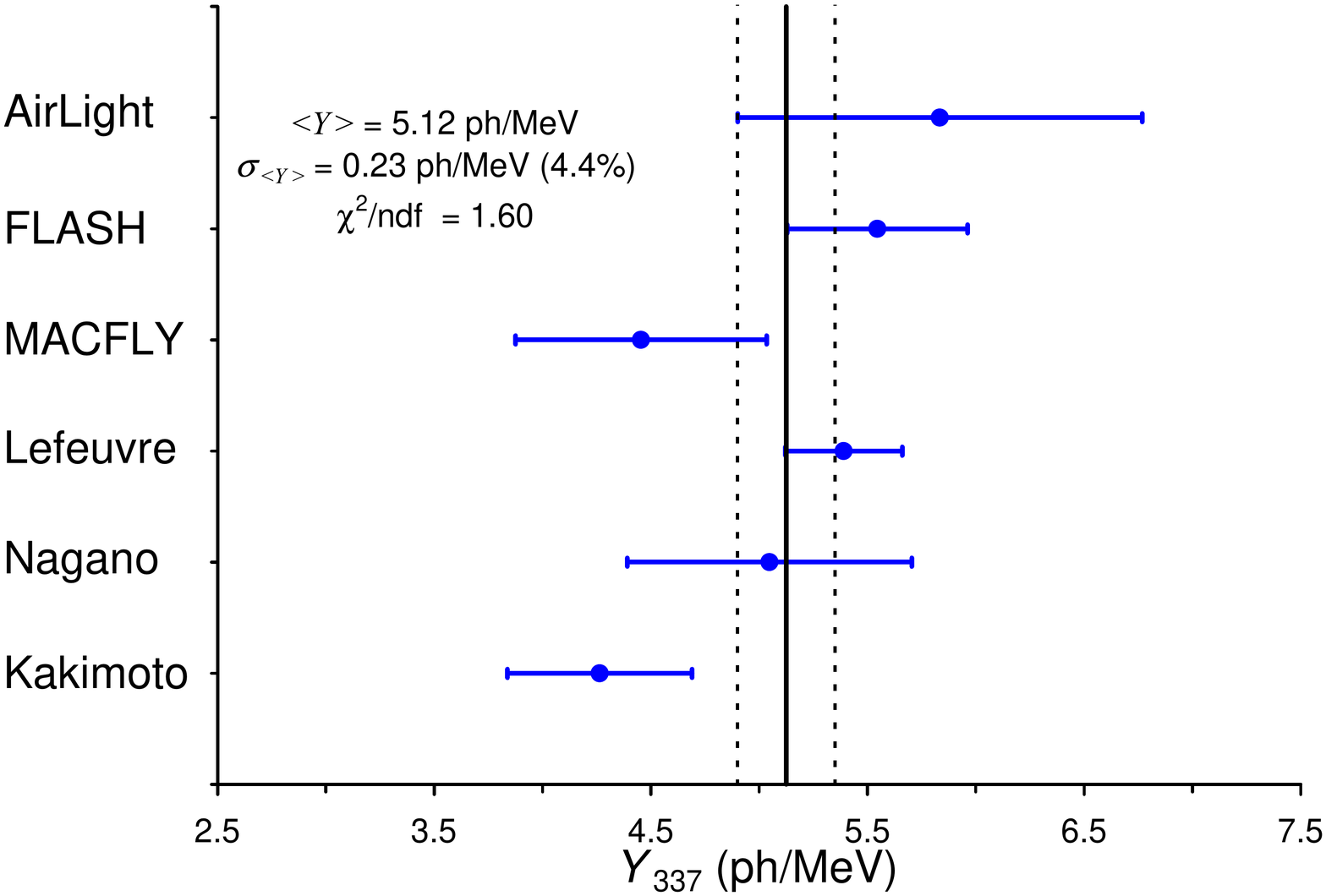}%
\label{fig:uncorr}%
}%

\subfloat[\footnotesize Corrected $Y_{337}$ values.]{%
\includegraphics[width=0.7\linewidth]{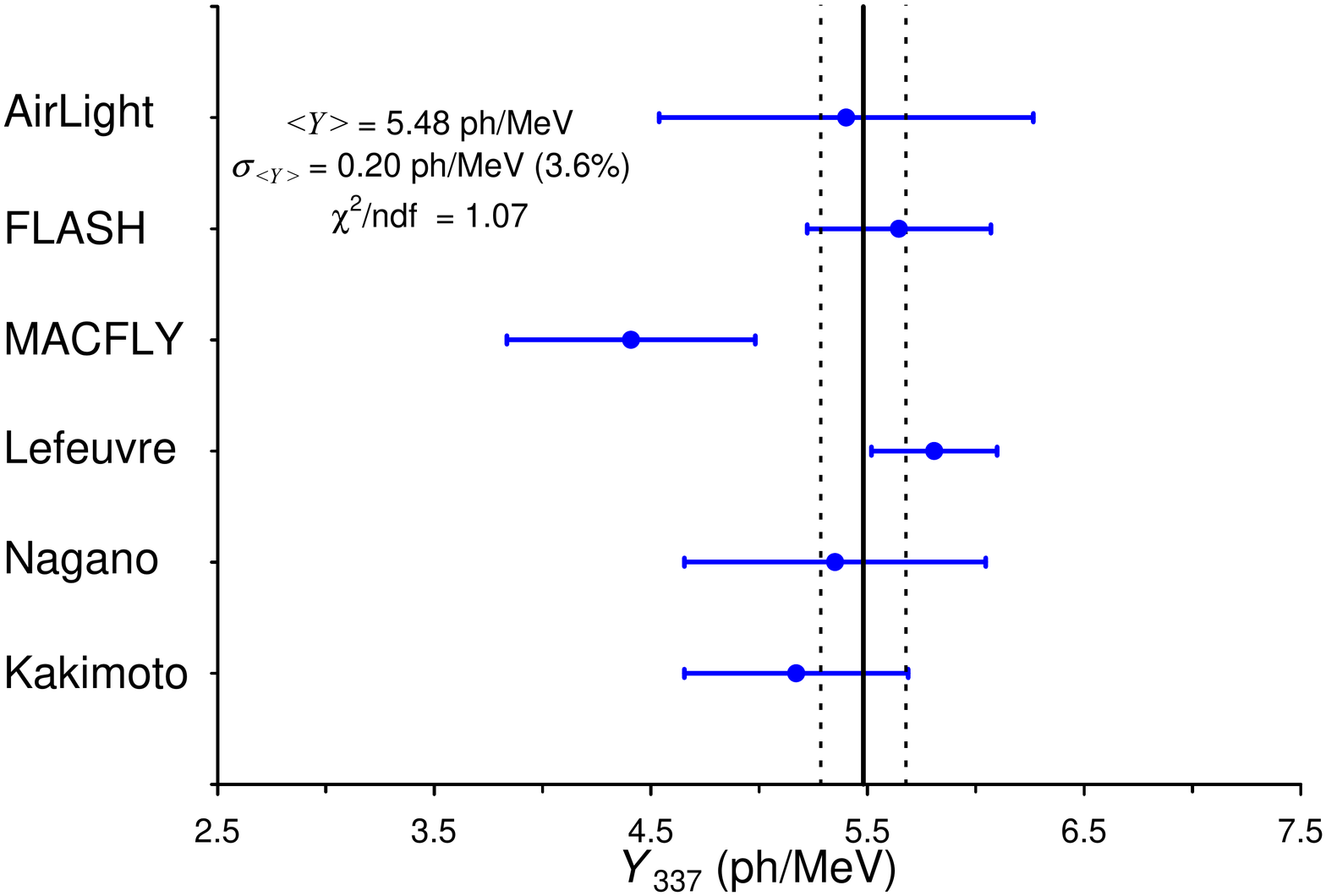}%
\label{fig:corr}%
}%
\centering\caption{\footnotesize Graphical representation of the $Y_{337}$ values at 1013~hPa and 293~K (see
table~\ref{tab:average}). The weighted mean value $\langle Y\rangle$ (vertical continuous line), the standard deviation
of the mean $\sigma_{\langle Y\rangle}$ (half of interval between dashed lines) and the $\chi^2$ statistic normalized
by the number of degrees of freedom are shown in the legends. (a)~Discrepancies between the uncorrected values are
sometimes larger than quoted errors, resulting in a relatively large $\chi^2/{\rm ndf}$ value. (b)~Better
agreement and consistency of the data sample are found when applying the proposed corrections.}%
\label{fig:average}
\end{figure}

Data listed in table~\ref{tab:average} for both the uncorrected and the corrected sample are plotted in
figure~\ref{fig:average}. The average value for the uncorrected $Y_{337}$ sample (figure~\ref{fig:uncorr}) is found to
be $\langle Y\rangle=5.12$~ph/MeV with $\sigma_{\langle Y\rangle}=0.23$~ph/MeV (4.4\%) and a $\chi^2/{\rm ndf}$ value
of 1.60, whereas the corresponding results for the corrected sample (figure~\ref{fig:corr}) are $\langle
Y\rangle$=5.48~ph/MeV, $\sigma_{\langle Y\rangle}=0.20$~ph/MeV (3.6\%) and $\chi^2/{\rm ndf}$=1.07. Therefore, our
corrections lead to a more consistent data sample suggesting that they do improve the determination of the deposited
energy for the different experiments.

In the absence of other objective criteria, data have been weighted using the quoted experimental uncertainties, as
mentioned above. Nevertheless, this weighting procedure has no significant effect on the final result. For instance,
assuming the same weight (i.e., the same variance) for all the experiments the results are $\langle
Y\rangle=5.09$~ph/MeV and $\sigma_{\langle Y\rangle}=0.25$~ph/MeV (5.0\% error) for the uncorrected data sample and
$\langle Y\rangle=5.30$~ph/MeV and $\sigma_{\langle Y\rangle}=0.20$~ph/MeV (3.8\% error) for the corrected one. The
weighted sample mean turns out to be somewhat larger than the non-weighted mean mainly due to the high statistical
significance of the results of Lefeuvre~\etal and FLASH, with the smallest associated uncertainties. In any case, a
similar decrease in the standard deviation is found when corrections are applied whichever weights are chosen.

As a further test, it has been checked that the weighted mean of the corrected data sample does not change
significantly if some measurement is excluded. For instance, removing the MACFLY result, which is the one showing the
largest departure from the average, would lead to 5.61~ph/MeV. Instead, excluding the result of Lefeuvre~\etal, with a
claimed uncertainty of 5\% that is incompatible with our correction of 7\%, an average fluorescence yield of
5.23~ph/MeV is obtained.

\begin{figure*}[t!]
\includegraphics[width=0.7\linewidth]{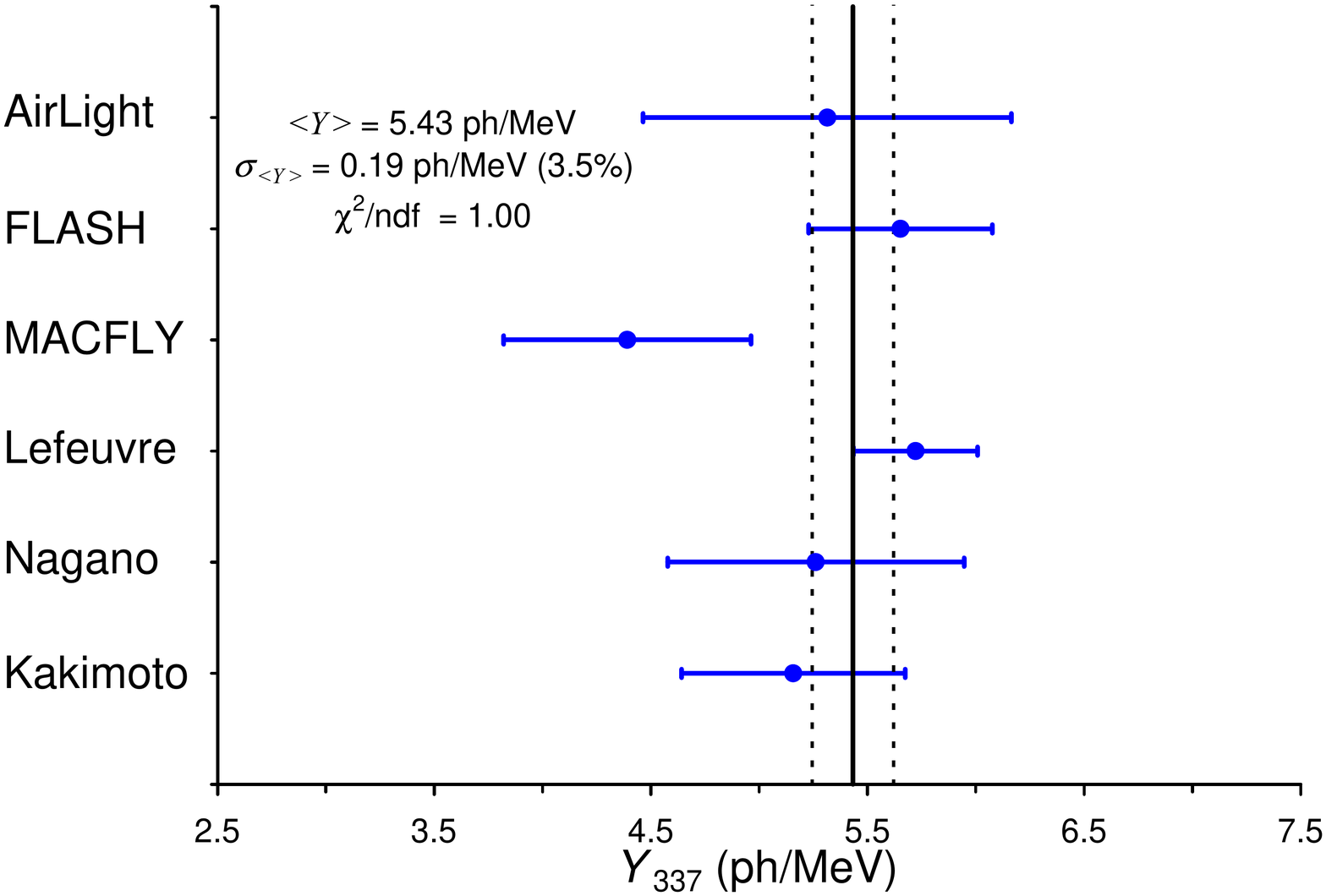}%
\centering\caption{\footnotesize The same as figure~\ref{fig:corr} but previously scaling to 100~MeV using the weak
energy dependence predicted by our simulation (see figure~\ref{fig:corr_E}). The consistency of the data sample is
strengthened slightly further.} \label{fig:corr_depE}
\end{figure*}

We have also studied the energy (in)dependence of the fluorescence yield in this data sample. Several experiments have
supported the assumed independence of the fluorescence yield at the level of about 10\% in several energy intervals. On
the other hand, the theoretical analysis described in~\cite{Arqueros_NJP} predicts a slight increase ($\sim 2\%$) with
decreasing energies in the $0.1-10$~MeV range, which is also compatible with experimental data (see
figure~\ref{fig:corr_E}). In principle, this effect should be included in order to quantify the consistency of the
available results. Average values listed in table~\ref{tab:average} have been recalculated by previously scaling all
the measurements to a common electron energy of 100~MeV according to the energy dependence predicted by our simulation.
The results are shown in figure~\ref{fig:corr_depE}. This energy scaling slightly lowers the $\chi^2/{\rm ndf}$ value
(from 1.07 down to 1.00) while both the average value and its uncertainty remain nearly unchanged (from $5.48\pm 0.20$
to $5.43\pm 0.19$~ph/MeV). Although this result would support the weak energy dependence predicted by our simulation,
we note that the evidence is still very weak and this small reduction of $\chi^2/{\rm ndf}$ after the energy correction
might have happened by chance. Therefore, even if confirmed experimentally\footnote{Currently not possible due to the
accuracy limitations in the determination of the energy dependence of the fluorescence yield in this large range.},
this weak energy dependence would not have any relevant impact on the energy reconstruction of cosmic rays using the
fluorescence technique.

Taking into account all the above considerations, including the possible effects of the weak energy dependence of the
fluorescence yield and the weighting procedure, a conservative result from our analysis is $Y_{337}=5.45$~ph/MeV with
an estimated error of $5\%$. According to the comparison of our simulation of the energy deposition with GEANT4, a
small systematic uncertainty of below 2\% should be added, although it does not affect the found improved compatibility
of results (i.e., the $\chi^2$ values). The recent absolute measurement of the AIRFLY collaboration~\cite{AIRFLY}
yields $Y_{337}=5.6$~ph/MeV with an uncertainty of $\lesssim5\%$ (still preliminary), which is fully compatible with
the above value. If this new result is included in the average, then a weighted mean value of 5.52~ph/MeV is obtained
from~(\ref{mean}) with an uncertainty of $\lesssim5\%$.

As already mentioned, for the comparison presented here we have normalized the air-fluorescence yield measurements to
its value for the 337~nm band at 1013~hPa and 293~K. However, in some occasions it might be more convenient to use the
integral of the fluorescence yield in a wider spectral range and/or other pressure and temperature conditions. The
conversion can be easily done following the procedure described in detail in~\cite{Rosado,Arqueros_NJP}. For instance,
the above average value would be of 20.1~ph/MeV ($\pm 5\%$) for the $300-420$~nm spectral range at the same reference
pressure and temperature, which would become 20.3~ph/MeV if the measurement of AIRFLY is included.

Our simulation can also provide a theoretical value of the air-fluorescence yield. Unfortunately, the evaluation of the
fluorescence emission cannot be very precise due to the large uncertainties in the relevant molecular parameters.
Therefore, we expect a large uncertainty in this calculation of the fluorescence yield, which we estimated to be about
25\%~\cite{Arqueros_NJP}. Nevertheless, a result\footnote{Using the quenching parameter of~\cite{AIRFLY_pressure}.} for
$Y_{337}$ of 6.3~ph/MeV has been found, which is consistent with the experimental ones, providing a valuable
theoretical support to these measurements.

In summary, the corrections to the measurements of the absolute air-fluorescence yield discussed in~\cite{Rosado}
increase significantly the compatibility of the various experimental results. An average value of $Y_{337}=5.45$~ph/MeV
with a 5\% uncertainty has been obtained. If the absolute fluorescence yield and error of AIRFLY are confirmed, a
consensus on this important parameter with an uncertainty below the 5\% level could be reached with high reliability.

\section*{Acknowledgements}

This work has been supported by the Spanish Ministerio de Ciencia e Innovacion (FPA2009-07772 and CONSOLIDER CPAN
CSD2007-42) and ``Comunidad de Madrid" (Ref.: 910600). J.~Rosado acknowledges a PhD grant from ``Universidad
Complutense de Madrid". The authors thank our colleagues of the Auger collaboration, in particular D.~Garc\'{i}a Pinto
and V.~Verzi, for fruitful discussions and comments on this work.

\end{document}